# Electrical contact properties between Yb and few-layer WS$_2$

*Shihao Ju, Lipeng Qiu, Jian Zhou, Binxi Liang, Wenfeng Wang, Taotao Li, Jian Chen, Xinran Wang, Yi Shi and Songlin Li*[a]

National Laboratory of Solid-State Microstructures, Collaborative Innovation Center of Advanced Microstructures, and School of Electronic Science and Engineering, Nanjing University, Nanjing, Jiangsu 210023, China

[a] Author to whom correspondence should be addressed: sli@nju.edu.cn

**ABSTRACT:** Charge injection mechanism from contact electrodes into two-dimensional (2D) dichalcogenides is an essential topic for exploiting electronics based on 2D channels, but remains not well understood. Here, low-work-function metal ytterbium (Yb) was employed as contacts for tungsten disulfide (WS$_2$) to understand the realistic injection mechanism. The contact properties in WS$_2$ with variable temperature ($T$) and channel thickness ($t_{ch}$) were synergetically characterized. It is found that the Yb/WS$_2$ interfaces exhibit a strong pinning effect between energy levels and a low contact resistance ($R_C$) value down to 5 kΩ·μm. Cryogenic electrical measurements reveal that $R_C$ exhibits weakly positive dependence on $T$ till 77 K, as well as a weakly negative correlation with $t_{ch}$. In contrast to the non-negligible $R_C$ values extracted, an unexpectedly low effective thermal injection barrier of 36 meV is estimated, indicating the presence of significant tunneling injection in subthreshold regime and the inapplicability of the pure thermionic emission model to estimate the height of injection barrier.

Two dimensional (2D) transition metal dichalcogenides (TMDs) with layered structure and chemical formula MX$_2$ (M = Mo, W, etc., X = S, Se, etc.) are considered as promising semiconductor channels for post-Moore microelectronics due to their fascinating properties.[1,2] According to the scaling rule of field-effect transistors (FETs),[3] the atomic channel thickness would represent the ultimately dimension that can avoid the short channel effect to the most and ensure the technological nodes beyond silicon. Among all TMDs, WS$_2$ is an outstanding member that features a low effective carrier mass,[4] suitable bandgaps[5] and good chemical stability, which makes it a strong candidate for high-performance electronics. However, WS$_2$ exhibits a high lying conduction band minimum (~0.35 eV higher than MoS$_2$), mainly determined by the high-energy 5d orbit of W atoms,[6] and hence it is more difficult for electrons to inject from electrodes to the W based TMD channels. Therefore, the inefficient charge injection between electrodes and WS$_2$ channels remains one of the main issues before practical applications.[7–9]

According to the Schottky-Mott rule,[10] $R_C$ originates from the presence of Schottky barriers at the electrode/channel interfaces. However, the physical scenario becomes more accurate after considering the interfacial states and the effect of Fermi level pinning (FLP), which largely reshape the alignment of energy levels and redefine the realistic barrier heights at interfaces. Such an FLP effect is ubiquitous in traditional bulk semiconductors such as silicon and germanium, as well as nominally dangling-bond-free 2D TMD semiconductors.[11–13] It is reported that the pinning factor $S \approx 0.11$ for 2D $MoS_2$,[14] suggesting a strong FLP effect and the presence of high-density interfacial states, which are introduced unintentionally in storage and during processing.

Numerous efforts have been devoted to improve the contact quality in FETs consisting of 2D TMD channels. One common strategy is inserting ultrathin insulating interlayers as tunneling media or stacking van der Waals contacts over TMD channels. The inserted interlayers, such as metal oxides and hBN,[15-22] between contacts and channels are expected to attenuate the metal-induced gap states and introduce interfacial dipoles, resulting in the reduction of FLP effect. Likewise, the role of stacked van der Waals contacts[23, 24] is to avoid perturbing the self-saturated surficial chemical bonds of 2D TMDs, which would minimize the density of interfacial states and alleviate the FLP effect as well. Another effective way to optimize contacts is performed through engineering the work function of contact metals ($\Phi_M$) According to the first-principle density functional theory,[25] the barrier heights are closely associated with $\Phi_M$ of electrodes, which has been confirmed in TMD FETs with contact electrodes such as Sc, Al and In.[14, 26, 27] Other contact optimization efforts, including chemically doping[28-30], phase engineering,[31] and edge contact,[32] have also been employed. Very recently, the semimetal Bi is reported to exhibit excellent contacts with monolayer TMDs; the reason is attributed to the low density and filling up of metal-induced gap states.[33] By and large, using an elemental metal as contacts without complicated manipulation still represents an attractive way for contact optimization.

Among all elemental metals, the lanthanide metals generally exhibit much lower $\Phi_M$ than Sc, Al and In. However, the use of the lanthanide metals as contacts to 2D TMDs remains elusive. In this work, we demonstrate that the lanthanide metal ytterbium (Yb) makes good contacts with few-layer $WS_2$, resulting in a low contact resistance ($R_C$) down to 5 kΩ·μm. Consistently, the cryogenic 4-probe measurement (4PM) and transfer length method (TLM) uncover weakly positive dependence of $R_C$ on temperature ($T$) till 77 K. Also, we reveal a negative correlation between $R_C$ and channel thickness ($t_{ch}$), which can be attributed to the increase of injection barrier due to the bandgap expansion from quantum confinement effect. The Yb electrodes exhibit a strong FLP effect, giving rise to comparable $R_C$ values between different 2D TMDs ($WS_2$ and $MoS_2$) and an unexpectedly low effective thermal injection barrier ($\Phi_B$) of 36 meV. The discrepancy in the magnitudes between $R_C$ and $\Phi_B$, together with its implication in the contact physics of 2D TMD channels, are also extensively discussed, in terms of the realistic $Yb/WS_2$ interfacial condition revealed by transmission electron microscopy (TEM).

Figure 1(a) illustrates the ideal band alignment for $WS_2$ and various metals with low and high work functions. In principle, the metal Yb would allow spontaneous electron injection into $WS_2$ if the FPL effect is absent or strongly suppressed, because the Fermi level of Yb lies higher than the conduction band minimum of $WS_2$. Figure 1(b) depicts the device structure of our $WS_2$ FETs, where mechanically exfoliated $WS_2$ flakes were first transferred onto predefined back dielectric/gate stacks and Yb/Ag/Au stacks were then thermally deposited as contacts. Hence, the FETs are Yb top contacted and $Al_2O_3$ or $SiO_2$ back gated. Detailed fabrication process is available in supplementary material. In Fig. 1(c), the interfacial condition between thermally deposited Yb and exfoliated $WS_2$ was characterized by TEM, which uncovers the damaged topmost $WS_2$ layer underneath Yb, a likely origin for the strong FLP effect and the $\Phi_B$ abnormality, as will be discussed later. Figure 1(d) shows the output curves under small bias, i.e., drain-source current ($I_{ds}$) versus drain voltage ($V_{ds}$), to check the contact quality in a typical Yb-contacted $WS_2$ FET [thickness ~ 2.4 nm, supplementary material Fig. S1(a)]. The linearity over a wide range of gate voltage ($V_{gs}$) from 5 to 30 V indicates the Ohmic nature of the Yb/$WS_2$ interfaces.

Figure 1(e) shows the transfer curves (i.e., $I_{ds}$ versus $V_{gs}$) at various $T$s. The on/off current ratios are higher than $10^8$ at all measured $T$s, along with a low off-state current ~0.1 pA at $V_{ds}$ = 1.1 V. The inset shows the transfer curves plotted in linear scale at 280 K at various $V_{ds}$s. In contrast to the ambipolar conduction behavior observed in Ti, Au and Pt contacted counterparts[21, 34, 35], the device is unipolar for n-type conduction merely because of the ultralow work function of Yb. Note that noticeable kinks emerge at the subthreshold region in the transfer curves, which can be ascribed to the contact turn-on effect caused by the $V_t$ variation between the different $WS_2$ areas underneath and outside contacts.[36] Additionally, we observe that the $I_{ds} - V_{gs}$ curves intersect at $V_{gs}$ ~ 45 V, which corresponds to an electron concentration ($n_{2D}$) of ~ $2.9 \times 10^{12}$ cm$^{-2}$. This point is associated with the transition from semiconductive to metallic behavior upon electronic gating, as commonly reported in high-quality TMDs.[37, 38] According to the Ioffe-Regel criterion,[39] metallic behavior emerges when $k_F \cdot l_e \approx 1$, where $k_F$ and $l_e$ are the Fermi wavenumber and electron mean free path, respectively. At the critical point, we estimate $k_F = \sqrt{2\pi n_{2D}} = 0.42$ nm$^{-1}$, $l_e = \hbar k_F \sigma / n_{2D} e^2 = 1.7$ nm and $k_F l_e$ = 0.74, which is in line with the theory.[37]

To shed light onto the magnitude of effective thermal $\Phi_B$ at the Yb/$WS_2$ interfaces, we performed variable-$T$ electrical characterization. The effective $\Phi_B$ of 2D FETs is normally extracted from the 2D thermionic emission (TE) equation.[15, 16, 22, 27, 33, 40, 41] Figure 1(f) plots the $\Phi_B - V_{gs}$ trend extracted from the Arrhenius relation [supplementary material Fig. S2 (c)]. For clarity, the diagrams of band bending for different transport regimes are also illustrated under the assumption of TE model. As commonly explained, the curve is linear in the TE regime until $V_{gs}$ exceeds the flat band voltage ($V_{fb}$). When $V_{gs} > V_{fb}$ the thermionic field emission (TFE, i.e., thermal assisted tunneling) mechanism gradually dominates, thus the curve deviates from linearity. Accordingly, we estimate an ultralow $\Phi_B$ of 36 meV and a $V_{fb}$ of 14 V. In comparison, the values of $\Phi_B$ are 120 and 200 meV reported in Ti and Pt contacted counterparts.[42, 43] The reduced $\Phi_B$ can be mainly attributed to the ultralow $\Phi_M$ of Yb contacts, which makes its Fermi level be pinned close to the conduction band minimum of $WS_2$. By plotting $\Phi_B$ versus $\Phi_M$ at various metal/$WS_2$ interfaces from literature,[42-47] we estimated $S$ ~ 0.13 [dashed blue line, Fig. 1(g)] without including our Yb data. The value is further lowered into the range of 0.01 ~ 0.06 by including Yb. The Yb data show a large deviation to the trend from literature. Such a $\Phi_B$ abnormality originates likely from the damaged topmost channel layer, as seen in the cross-sectional TEM image in Fig. 1(c), which acts equivalently as an ultrathin insulating interlayer to lower $\Phi_B$ and increases the portion of tunneling current for charge injection. In this sense, the injection barrier $\Phi_B$ extracted with the commonly adopted

TE model loses its original physical meaning, which, in this case, represents only an effective quantity of thermal barrier.

To gain insight into the practical contact quality, we further double-checked the $R_C$ values by two independent approaches, including TLM and 4PM. Here the WS$_2$ FETs are all fabricated on substrates with 30 nm Al$_2$O$_3$ as dielectric for high coupling capacitance. We first extracted $R_C$ by applying TLM on multiple FETs with varied channel lengths ($L_{ch}$). Figure 2(a) shows the optical image of 4 FETs with $L_{ch}$ ranging from 2.3 to 5.1 µm, where the channel areas are labelled by red dotted lines. We note that the threshold voltage ($V_T$) is prone to change as $L_{ch}$ varies. Hence, it is more appropriate to compare $R_C$ at a fixed $n_{2D}$, which is proportional to $V_{gs} - V_T$. Figure 2(b) plots the width-normalized total resistance $R_{Total}$ versus $L_{ch}$ at three typical $n_{2D}$ values and Fig. 2(c) summarizes the $R_C$ values at different $T$s. We estimate a low $R_C$ value of ~13.1 kΩ·µm at 300 K at $n_{2D} = 10^{13}$ cm$^{-2}$, which is one order lower in magnitude than those in conventional Ti and Ni contacted counterparts.[28, 48] At cryogenic $T$s, we observe a weakly positive correlation between $R_C$ and $T$ till 77 K.

The values of $R_C$ were then jointly evaluated by the standard 4PM. In order to increase the accuracy in evaluation, here only the flake portions with regular shapes were used as FET channels and the two inner voltage electrodes were defined with sufficiently large ratios of distance to width. Figure 2(d) shows the optical image for a typical FET with the 4-probe geometry. First, we calculate and compare the 4-probe ($\mu_{4p}$) and 2-probe mobility ($\mu_{2p}$) at different gating conditions. As shown in Fig. 2(e), $\mu_{2p}$ exhibits a same trend to $\mu_{4p}$ in a wide range of gating conditions where their values are consistent. Despite the scanning hysteresis, they reach the peak mobility of ~ 60 (~70) cm$^2$ V$^{-1}$ s$^{-1}$ in the forward (backward) scanning direction. The duplication between $\mu_{2p}$ and $\mu_{4p}$ indicates that there is a similar gate dependence between $R_C$ and the sheet resistance of channels ($R_S$), which is often inappropriately explained as negligence of $R_C$ in literature. Figure 2(f) displays the estimated $R_C$ versus $n_{2D}$ at $T$ of 77, 150, and 300 K. Two noticeable features are seen: 1) $R_C$ shows a low value of ~ 7.3 kΩ·µm at 300 K, comparable with the TLM data above (13.1 kΩ·µm); 2) $R_C$ also exhibit a weakly positive correlation with $T$ at fixed $n_{2D}$, after considering the uncertainty of $n_{2D}$ values arising from $V_T$ extraction [supplementary material Fig. S6(a)].[9]

In principle, $R_C$ is governed by the current injection equation $R_C = \sqrt{\rho_C R_S}\coth(L_C/\sqrt{\rho_C/R_S})$,[49] where $\rho_C$ and $L_C$ denotes the specific contact resistivity and the physical contact length, respectively, which plus $R_S$ are three fundamental device quantities. In our widely contacted devices with $L_C \geq 1$ µm, $\coth(L_C/\sqrt{\rho_C/R_S}) \sim 1$ and $R_C \sim \sqrt{\rho_C R_S}$. Hence, the trend of $R_C$ variation with different parameters (e.g., $T$ or $n_{2D}$) can be understood in terms of the changes of $\rho_C$ and $R_S$. Since $\rho_C$ and $R_S$ are both negatively correlated with $n_{2D}$, one can quickly understand the overall negative correlation of $R_C$ with $n_{2D}$, as shown in Figs. 2(c) and 2(f).

In contrast, the $R_C - T$ relation becomes more complicated because it is determined by the competition between two opposing trends between $\rho_C$ and $R_S$ with $T$. For high-quality TMD channels, $R_S$ is constantly correlated positively to $T$ because of the suppressed phonon scattering at cryogenic $T$s [supplementary material Figs. S4(c) and 4(d)]. However, $\rho_C$ generally depends negatively on $T$, but the degree of dependence is closely associated with the dominant injection mechanism. With $n_{2D}$ changes from low to high levels by electronic gating, the charge injection mechanism evolves from TE, TFE, to direct field emission (FE) and results in a transition from strong to weak dependence on $T$. Also, the presence of the damaged TMD layer between channels and electrodes would increase the component of tunneling and further reduce the $T$ dependence [Supplementary material Figs. S4(b) and 4(e)]. The overall weak $R_C - T$ correlation dependence indicates the non-dominancy of TE mechanism even at the low $n_{2D}$ regime, which predicts a strong $T$ dependence as $\rho_C \propto T^{-3/2} \exp\left(\frac{q\phi_B}{k_B T}\right)$. In this regard, the extracted value of thermal $\Phi_B$ represent merely an effective quantity for most 2D TMDs. This argument becomes more reasonable when considering the presence of the damaged $WS_2$ layers underneath contacts [Figs. 1(c)].

In addition to $WS_2$, we also prepared and characterized the electronic properties of the Yb contacted $MoS_2$ FETs, whose cryogenic transfer characteristics are given in Fig. 3(a). Similar to $WS_2$, the $MoS_2$ device shows n type behavior, linear output characteristics, and a weak $R_C - T$ correlation [supplementary material Fig. S6(c)]. In analogue to $WS_2$, $R_C$ of the $MoS_2$ FET with a same $t_{ch}$ is reduced from ~300 to ~7 kΩ·μm as $n_{2D}$ increases from 1 to $8 \times 10^{12}$ cm$^{-2}$, as seen in Fig. 3(b). No remarkable difference is observed in $R_C$ between the $WS_2$ and $MoS_2$ channels within the experimental uncertainty, corroborating again the strong PFL effect of Yb contacts to TMD channels.

To fully understand the effect of channel thickness on $R_C$, the contact properties were also collected for $WS_2$ FETs with different $t_{ch}$ values [Fig. 3(c)]. The general tendency is $R_C$ reduces as $t_{ch}$ increases from 3 to 7 layers, where it reduces from 70 to 15 kΩ·μm at mediate $n_{2D} = 4 \times 10^{12}$ cm$^{-2}$ and reduces from 37 to 6 kΩ·μm at high $n_{2D} = 8 \times 10^{12}$ cm$^{-2}$. The negative $R_C$-$t_{ch}$ correlation is consistent with the trends of $\rho_C$ and $R_S$, because, as $t_{ch}$ reduces, the charge mobility generally decreases[50] and the injection barrier $\Phi_B$ increases.[7] We also collected the contact property of Yb with the defective 1L CVD grown $WS_2$ (supplementary material Fig. S7), in which $\Phi_B$ (i.e., $\rho_C$) is maximized due to larger bandgap and $R_S$ is higher due to the presence of large densities of defects (i.e., increased charge scattering). In this case, $R_C$ is significantly degraded by 3–4 orders in magnitude, as compared with the high-quality few-layer channels. Hence, the $R_C$ magnitude is also strongly associated with the channel quality and can be improved by adopting high-quality samples.[51]

Finally, in Fig. 3(d) we compare the contact characteristics by summarizing the $R_C$ values of $WS_2$ with various metals.[9, 27-30, 32, 38, 42, 48, 51] Despite the damage to the topmost channel layer and the presence of strong FLP effect, the Yb metal still represents one of the few elemental materials resulting in $R_C$ below 10 kΩ·μm. At high $n_{2D}$, $R_C$ reaches down to 5 kΩ·μm for the 6L $WS_2$. In contrast, the metal contacts with medium $\Phi_M$ (Ag, Ti, Ni, Au, graphene and Cr) tend to result in high $\Phi_B$ and $R_C$. It is expected that the contact performance can be further optimized if the damaging effects to the topmost channel layer are suppressed.

In conclusion, we performed thorough evaluations on the contact quality between Yb and $WS_2$ with variable thickness via cryogenic TLM and 4PM. Thanks to the ultralow work function of Yb, the Fermi level of contacts lies close to the conduction band minimum of $WS_2$, resulting in an effective thermal barrier of 36 meV and a low contact resistance down to 5 kΩ·µm. Also, we identified the presence of damaged topmost channel layer and observed a weakly positive $R_C - T$ correlation, suggesting that the thermal assisted tunneling current, rather than the pure thermionic emission, is the dominant charge injection mechanism, even in the subthreshold regime in the 2D TMD channels. Remarkably, the contact quality of Yb outperforms most common metals. The results show the potential of contact engineering for 2D TMDs in ultrathin-body electronics by utilizing the low-work-function lanthanide metals as contact electrodes.

## SUPPLEMENTARY MATERIAL

See the supplementary material for the synthesis of $WS_2$ crystals and monolayer $WS_2$, deposition of $Al_2O_3$, device fabrication and characterization of $WS_2$ FETs, the extraction of Schottky barrier heights, output curves at low temperatures and more detailed electrical measurements for $WS_2$ and $MoS_2$.

## ACKNOWLEDGMENTS

This work was supported by the National Natural Science Foundation of China (61974060, 61674080 and 61521001), the National Key R&D Program of China (2021YFA1202903), the Innovation and Entrepreneurship Program of Jiangsu province and the Micro Fabrication and Integration Technology Center in Nanjing University.

## AUTHOR DECLARATIONS

**Conflict of Interest**

The authors declare no conflicts of interests.

## DATA AVAILABILITY

The data that support the findings of this study are available from the corresponding author upon reasonable request.

**Figure captions**

FIG. 1. (a) Ideal band alignments for various contact metals and $WS_2$. (b) Schematic diagram for Yb contacted $WS_2$ FETs. (c) Interfacial condition between Yb and $WS_2$ revealed by TEM. The dotted and solid golden lines denote the damaged and intact $WS_2$ layers, respectively. (d) Linear output curves at various $V_s$. (Inset: Optical image of the device) and (e) transfer characteristics for a typical Yb-$WS_2$ FET at cryogenic $T$s. Inset: Linear plots for transfer curves. (f) Effective $\Phi_B$ versus $V_{gs}$ and corresponding band diagrams. (g) Comparison of $\Phi_B$ with different metal contacts. $t_{ch}$ values of TMDs are given after the sign "+". The dashed blue line is the fit without including Yb. The two dashed black lines represent the upper and lower limits fitted with Yb. Except Yb, all the other $\Phi_B$ values are from Refs. 42-47.

FIG. 2. $R_C$ extraction with TLM and 4PM. (a) Optical image for typical multiple FETs with varied $L_{ch}$. The $WS_2$ channels are labelled by dotted red lines. (b) $R_{Total}$ versus $L_{ch}$ at different $n_{2D}$. (c) $R_C$ versus $n_{2D}$ extracted with TLM. (d) Optical image for a typical FET with four-terminal geometry. Inset: Electrode distances in unit of μm. (e) Comparison of $\mu_{2p}$ and $\mu_{4p}$. (f) Extracted $R_C$ versus $n_{2D}$ by 4PM. The error bars of horizontal axis in (c) and (f) originate from the uncertainty of $V_T$.

FIG. 3. (a) Transfer characteristics for a typical Yb-$MoS_2$ FET. Inset: Optical image of the device. (b) Extracted $R_C$ versus $n_{2D}$ for $WS_2$ and $MoS_2$. (c) $R_C$ versus $t_{ch}$ of $WS_2$. (d) Comparison of $R_C$ of $WS_2$ among different contact materials.

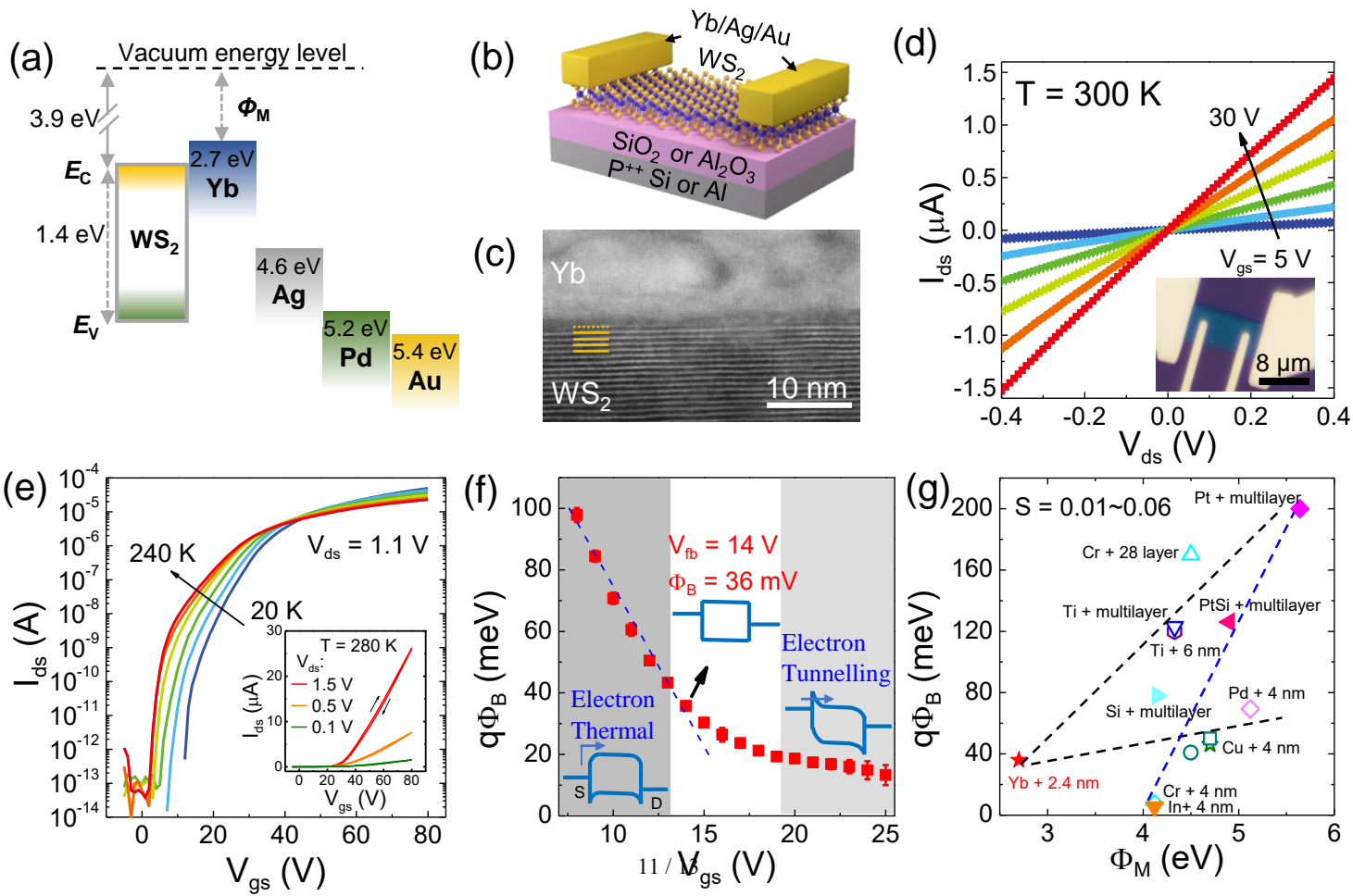

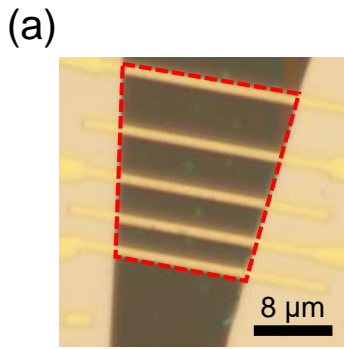 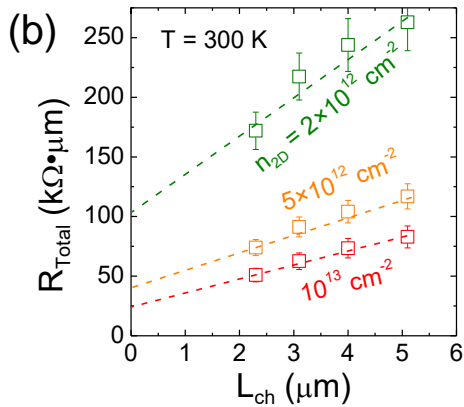 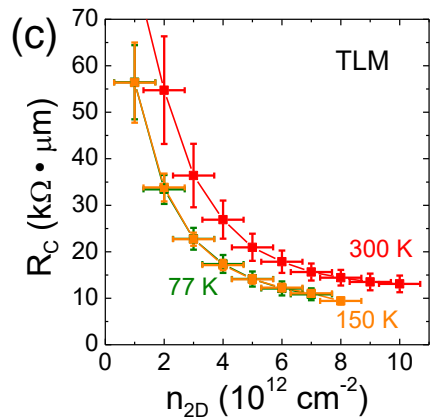
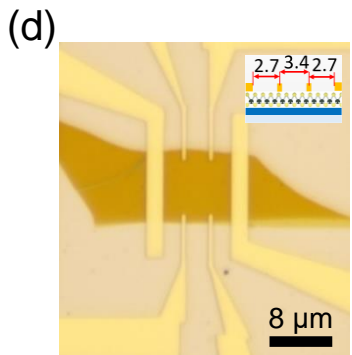 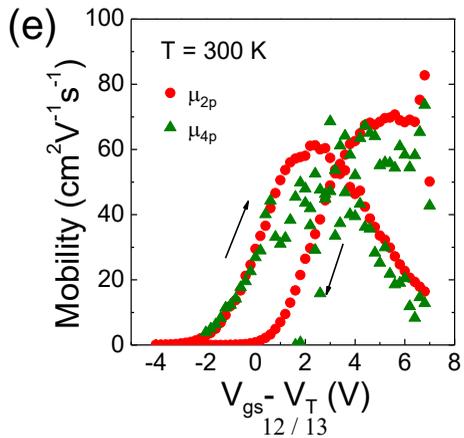 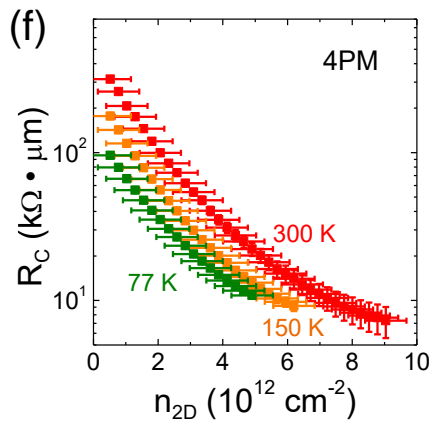

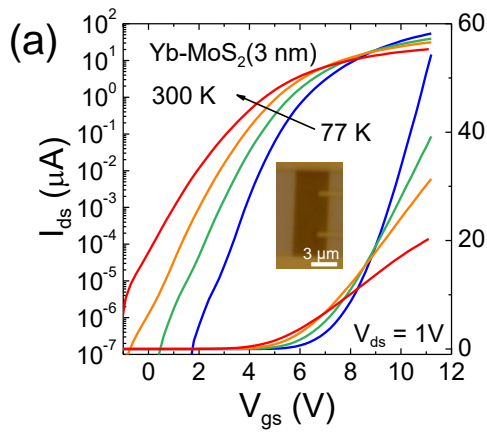
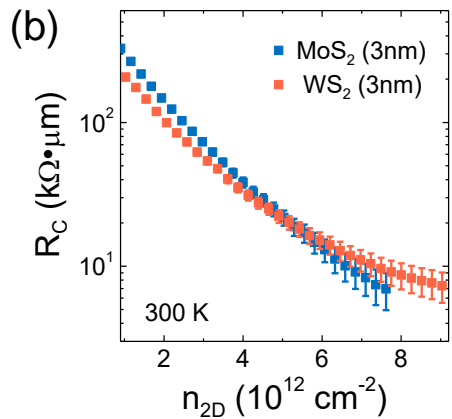
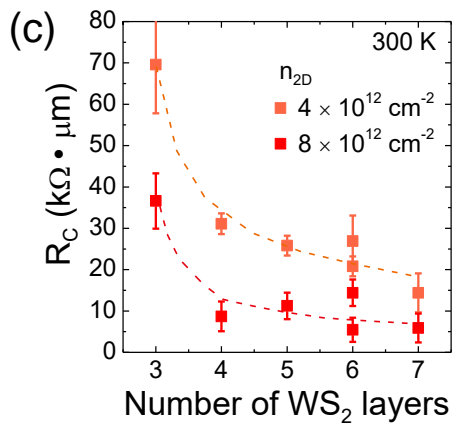
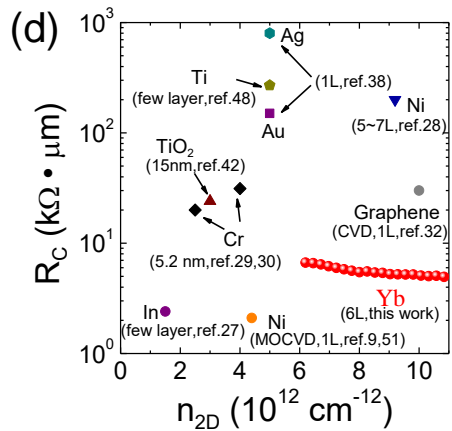

# Electrical contact properties between Yb and few-layer WS$_2$

*Shihao Ju, Lipeng Qiu, Jian Zhou, Binxi Liang, Wenfeng Wang, Taotao Li, Jian Chen*[a]*, Xinran Wang, Yi Shi and Songlin Li*[a]

National Laboratory of Solid-State Microstructures, Collaborative Innovation Center of Advanced Microstructures, and School of Electronic Science and Engineering, Nanjing University, Nanjing, Jiangsu 210023, China

[a] Authors to whom correspondence should be addressed: chenj63@nju.edu.cn or sli@nju.edu.cn

**Synthesis of WS$_2$ crystals**

Bulk WS$_2$ crystals were grown by a chemical vapor transport method. A quartz tube containing a mixture of WS$_2$ (99.9%, 3A Chemicals) and iodine powders were heated and sealed as an ampoule at a pressure of ~ $10^{-3}$ Pa and was then placed into a two-zone furnace. Afterwards, the mixture was heated with a rate of 9.5 K per minute from 300 K to 1320 K and then WS$_2$ was slowly sublimated and transported with iodine vapor from the high- to low-temperature zone. After 168 hours of growth, the quartz tube was cooled to room temperature within 24 hours. Finally, shiny and hexagonal WS$_2$ crystals were collected at the low-temperature zone of the ampoules.

**Atomic layer deposition of Al$_2$O$_3$**

The Al$_2$O$_3$ dielectric was deposited at 150 °C by an atomic layer deposition system. The precursors (trimethyl-aluminum and H$_2$O) were repeatedly carried into the reaction chamber via nitrogen flow (20 sccm). The pulse time for trimethyl-aluminum and $H_2O$ was both 30 ms and the purge time was 30 s between pulses. The growth rate of $Al_2O_3$ was about 1 angstrom/cycle with the reaction conditions.

**Device fabrication and electrical characterization**

Multilayer $WS_2$ ($MoS_2$) devices: The 2D $WS_2$ ($MoS_2$) flakes were mechanically exfoliated from bulk crystals with Scotch tape in a nitrogen filled chamber in a glovebox system with the levels of $O_2$ and $H_2O$ less than 1 ppm. Then the exfoliated flakes were directly transferred, assisted by polydimethylsiloxane films, onto the 90 nm $SiO_2$ capped highly doped Si or onto 30 nm ALD $Al_2O_3$ capped aluminum electrodes. Next, the samples were delivered to another chemical chamber of the glovebox to spin coat the bilayer electron resists: methyl methacrylate (MMA, EL9) and poly(methyl methacrylate) (PMMA, A4). After patterned with source and drain contact regions via electron beam lithography, the resists were then developed for 40 s in ambient surrounding and immediately carried into a thermal evaporation chamber for metallization. We purposely deposited 10 nm Yb at a high rate of ~ 1 Å/s to alleviate the rapid metal oxidation during evaporation, followed by evaporating the capping layers of 60 nm Ag and 10 nm Au. After metal lift-off in cold acetone and rinsed in isopropanol, the samples were finally taken into the probe station for electrical measurements.

Monolayer $WS_2$ devices: Monolayer $WS_2$ films grown on sapphire substrates were transferred to the target substrates using a PMMA assisted wet transfer process. First, the sapphire substrate with monolayer $WS_2$ films was spin coated with PMMA A4. The corners of PMMA films were scratched via a razor blade and a piece of thermal release tape was attached on the PMMA surface. The substrate was then immersed in a NaOH saturated solution kept at 60 °C. The NaOH solution invaded into the sapphire/$WS_2$ interface and separated the thermal release tape/PMMA/$WS_2$ film from the substrate. The separated film was rinsed with deionized water for three times and gently attached to the target 30 nm $Al_2O_3$ substrate. Finally, the substrate was baked on the hot plate to remove the thermal release tape and moisture and then immersed in acetone to remove the PMMA layer. The monolayer WS$_2$ film was patterned by electron beam lithography and then etched by CF$_4$ in reactive ion etching to form the rectangle channel.

All the electrical measurements were performed with Keithley 2636B sourcemeters under vacuum (air pressure <1×10$^{-3}$ Pa) in a cryogenic probe station (CRX-6.5K, Lakeshore).

**Extraction of the Schottky barrier height**

The thermal $\Phi_\text{B}$ can be extracted from the 2D thermionic emission equation which describes the current across a Schottky junction into 2D semiconductors at the subthreshold regime:

$$I_\text{ds} = A_\text{2D}^* A T^{\frac{3}{2}} \exp\left[-\frac{q}{k_\text{B}T}\left(\Phi_\text{B} - \frac{V_\text{ds}}{n}\right)\right]$$

where $A_\text{2D}^*$ is the 2D Richardson constant, $A$ is the junction area, $k_\text{B}$ is the Boltzmann's constant and $n$ is the ideality factor accounting for barrier lowering by image charges. With this formula, the Arrhenius plots, i.e. $\ln(I_\text{ds}T^{-3/2})$ against $1000/T$, at 14 V gate bias are plotted in Fig. S1(c) under various $V_\text{ds}$. Then, the slopes of Arrhenius plots are further extracted by linear fitting and summarized in Fig. S1(d) to determine $\Phi_\text{B}$ via the relation $S(V_\text{ds}) = -1000q(\Phi_\text{B} - V_\text{ds}/n)/k_\text{B}$. As marked in Fig. S1(d), the intercept at zero $V_\text{ds}$ ($S_0$) extracted by linear fitting $S(V_\text{ds})$ against $V_\text{ds}$ yields the values of $\Phi_\text{B}$ at different $V_\text{gs}$.

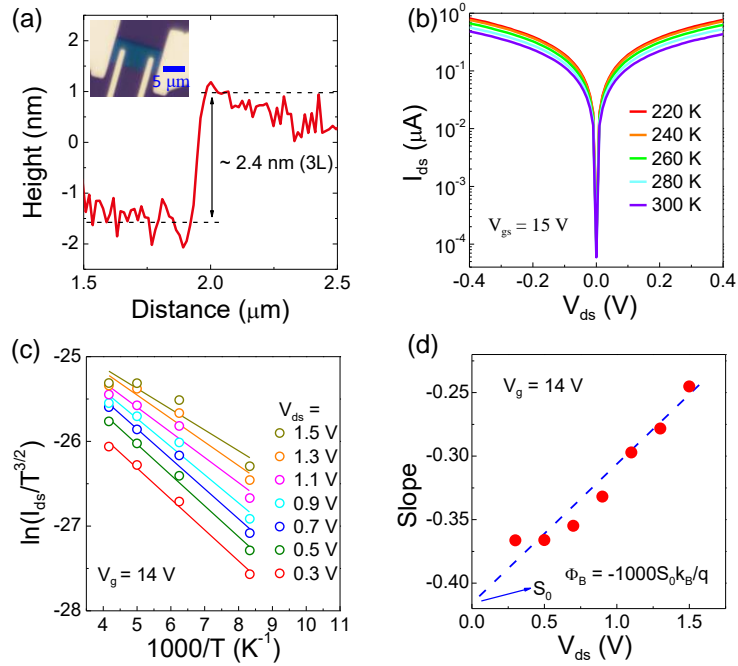

FIG. S1. (a) The optical image (inset) and AFM characterization of a typical $WS_2$ sample on 90 nm $SiO_2$. (b) $I_{ds} - V_{ds}$ characteristics at varied $T$ from 220 to 300 K near the flat band voltage ($V_{fb} \sim 14$ V). (c) Arrhenius plot for $\ln(I_{ds}/T^{3/2})$ versus 1000/T with variable $V_{ds}$. (d) Corresponding slopes extracted from (c) under different $V_{ds}$ values, where the Schottky barrier height can be extracted from the intercept of the y-axis at $V_{ds} = 0$.

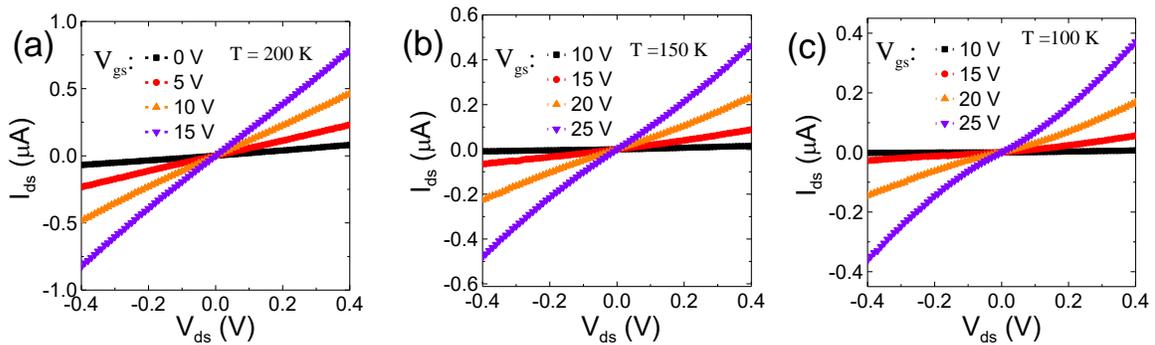

FIG. S2. (a) - (c) Evolution of the linearity with temperature for the low-temperature $I_{ds} - V_{ds}$ output characteristics measured at gate voltages within the subthreshold region from 200 to 100 K.
x

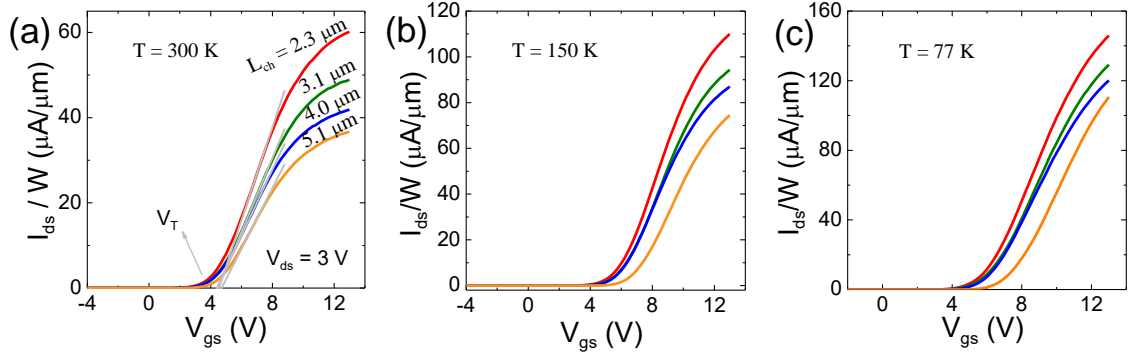

FIG. S3. TLM measurements for the Yb contacted $WS_2$ FETs on substrates with 30 nm $Al_2O_3$ as dielectric. Corresponding transfer curves for the FETs at different temperatures of (a) 300 K, (b) 150 K and (c) 77 K.

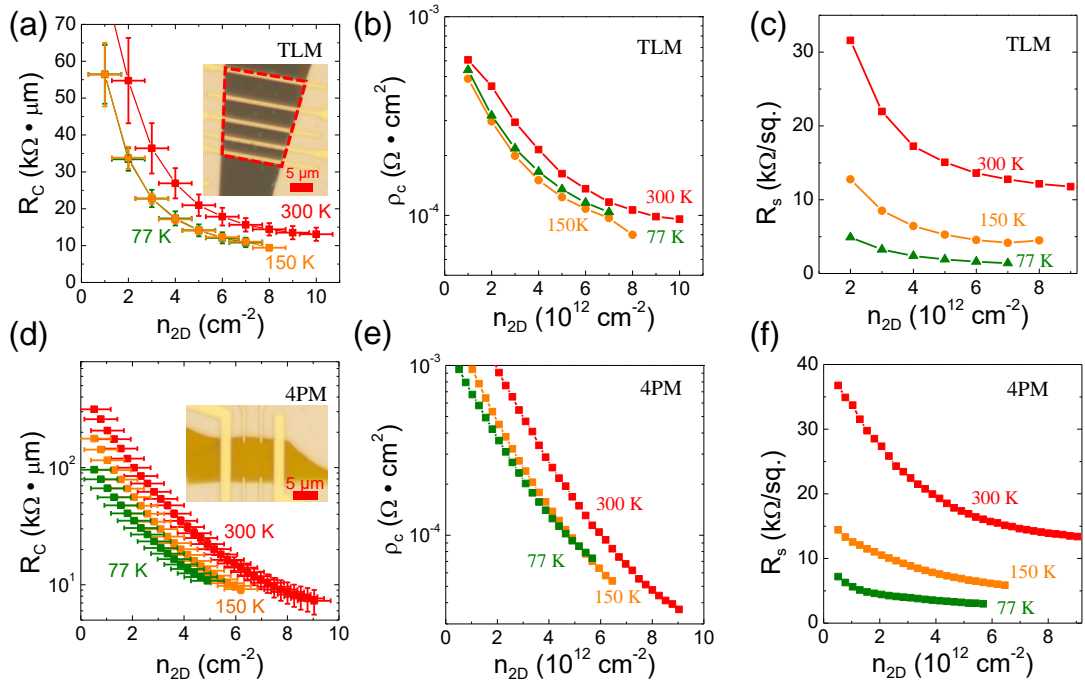

FIG. S4. Temperature dependent $\rho_C$ and $R_S$ extraction. Extracted $R_C$ versus $n_{2D}$ by (a) TLM and (b) 4PM. Extracted $n_{2D}$ dependent $\rho_C$ via (b) TLM and (e) 4PM. Corresponding $R_S$ versus $n_{2D}$ of the (c) TLM and (f) 4PM samples.

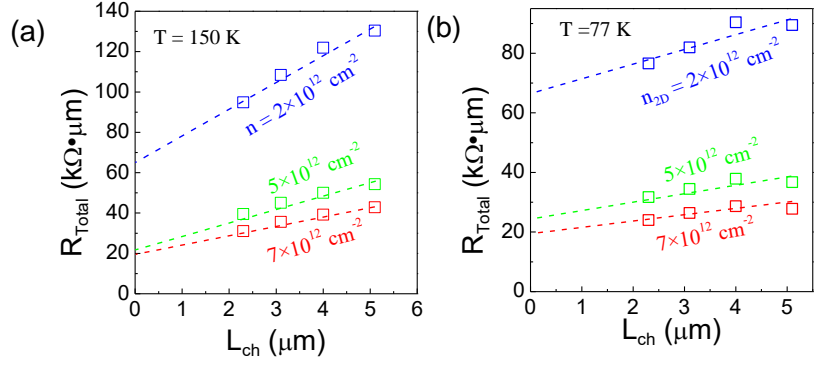

FIG. S5. (a) and (b) Width-normalized total resistance $R_{Toatl}$ for specific carrier concentration at 150 and 77 K, respectively.

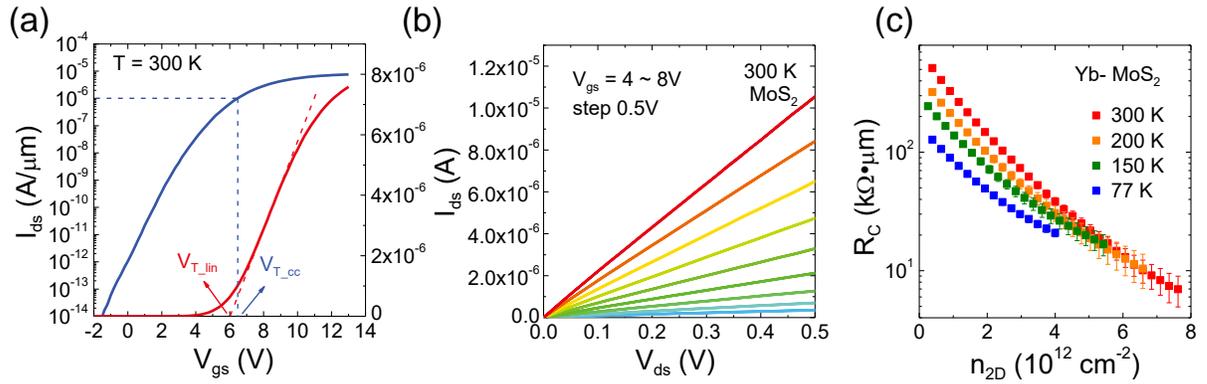

FIG. S6. (a) Illustration of extraction of $V_T$ by linear extrapolation ($V_{T\_lin}$) of the transfer characteristics and constant-current method ($V_{T\_cc}$). (b) The output curves of a Yb-contacted MoS$_2$ FET. (c) Extracted $R_C$ versus $n_{2D}$ at 77, 150, 200 and 300 K for the MoS$_2$ device.

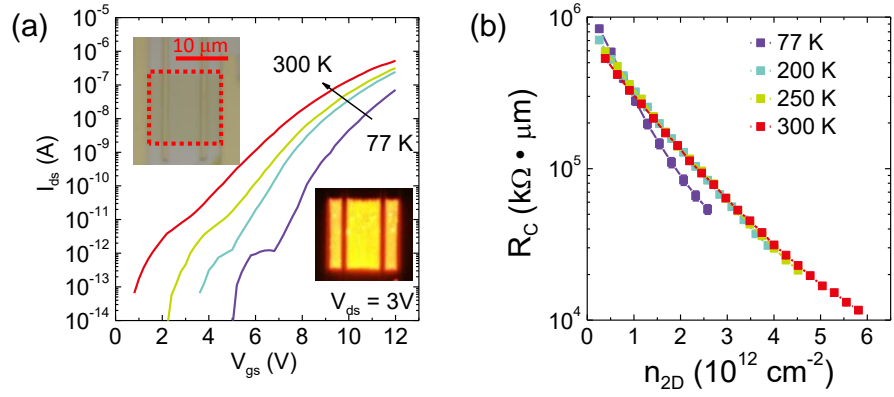

FIG. S7. (a) Transfer characteristics for a typical Yb contacted defective 1L CVD WS$_2$ FET on 30 nm Al$_2$O$_3$ at varied $T$s; insets: optical and PL images of the device. (b) $n_{2D}$ dependent $R_C$ for the 1L WS$_2$ FET.